\documentclass[preprint,preprintnumbers,amsmath,amssymb,floatfix,showpacs]{revtex4}
\usepackage{epsfig}

\begin{document}
\title{Is the Wannier threshold law for angular distribution in double photoionization of
atoms true at practically attainable energies of ejected electrons?}
\author{Vladislav V. Serov and Tatyana A. Sergeeva}
\affiliation{Chair of Theoretical and Nuclear Physics, Saratov
State University, 83 Astrakhanskaya, Saratov 410026, Russia}

\date{\today}

\begin{abstract}
We calculated {\it ab initio} the three-fold differential cross section of a double single-photon Helium photoionization at the equal energy sharing, and obtained from one the Gaussian width parameter $\gamma$, describing the angular interelectron correlations, for the total electrons energies range $E$ from 0.1 eV to 100 eV. The results are in the excellent agreement with experimental data but indicate that the Wannier threshold law for the angular distribution $\gamma\propto E^{1/4}$ is not correct at energies attainable in modern experiments. It is shown that the $\gamma$ dependence on the energy is much better described by the modified threshold law, obtained by Kazansky and Ostrovsky [J. Phys. B \textbf{26}, 2231 (1993)]. 
Also, we explored the Gaussian width parameter for a double photoionization of the targets with the strongly asymmetrical initial state configuration: the atomic Hydrogen negative ion H$^-$ and the Helium in the $2s\,^1$S and $3s\,^1$S excited states. We found that the Gaussian width dependence on the total ejected electrons energy for these targets has a maximum at low energies. We show also that the correlation parameter dependence on the interelectron angle for these targets is essentially non-Gaussian and has a number of peaks equal to a number of initial state radial nodes, that reveals the new abilities for the qualitative analysis of the electron structure. 
\end{abstract}


\maketitle

\section{Introduction}
It is known, that if the incident radiation is linearly
polarized in the $Oz$ direction, the three-fold differential cross section (3DCS) of a double photoionization of an atom by a single photon can be represented via
gerade and ungerade amplitudes \cite{Huetz1991}
\begin{eqnarray}
\frac{d^{3}\sigma}{dE_1d\Omega_1d\Omega_2}=
|a_g(E_1,E_2,\theta_{12})(\cos\theta_{1}+\cos\theta_{2})
 +
 a_u(E_1,E_2,\theta_{12})(\cos\theta_{1}-\cos\theta_{2})|^2,\label{agau}
\end{eqnarray}
where the ungerade amplitude $a_u=0$ for electrons energies $E_1=E_2$, and the gerade
amplitude usually called a correlation parameter can be approximated at low electrons energies by a Gaussian curve 
\begin{eqnarray}
a_g(E_1, E_2, \theta_{12})\simeq A\exp\left[-2\ln
2\frac{(\theta_{12}-\pi)^2}{\gamma^2}\right],\label{GaussianAppr}
\end{eqnarray}
as it was shown by Rau \cite{Rau1976} following WannierТs theory \cite{Wannier1953}.
The Gaussian width parameter $\gamma$ is a single angle parameter describing the angular distribution, therefore it is often used for the analysis of strength of the angular interelectron correlation. When $\gamma$ is large then the interelectron correlation is weak and vice versa. According to the Wannier threshold law, near the double photoionization threshold it should be 
\begin{eqnarray}
\gamma=\gamma_0 E^{1/4} \label{WannierLaw}
\end{eqnarray}
where $E=E_1+E_2$ is a total ejected electrons energy. In spite of the fact that the energy range of the correctness of this law is not established yet, experimentators and theorists still use the expression (\ref{WannierLaw}) for the data interpretation, trying to find the scaled
width parameter $\gamma_0$. A lot of formulas for $\gamma_0$ have been proposed by various authors \cite{Otranto2005}. However, Kazansky and Ostrovsky have shown \cite{Ostrovsky1993}, that the Wannier threshold law (\ref{WannierLaw}) is correct only when the electrons deceleration by the nucleus field is neglected or for extremely law energies of the order of $10^{-5}$ eV. From the other hand, we has found in \cite{Serov2008} that $\gamma$ for the Hydrogen negative ion H$^-$ starts to grow with energy decreasing at low energies in the obvious contradiction with the Wannier threshold law. The aim of the present work is a calculation of the Gaussian width $\gamma$ at the smaller energies.

\section{Details of the calculation procedure}
In our calculations we used the time-dependent scaling (TDS) method \cite{Serov2007,Serov2008}. The main advantage of this method is the 3DCS obtaining for all ingoing photon energy values by the one act of computation. The problem is the presence of not only the wave function component corresponding to the double ionization and being described correctly at the extending coordinate system, but also components describing the bound states and the single ionization states which are described incorrectly for the large values of the expansion coefficient $a(t)$ (see \cite{Serov2007}). Therefore at evolution times $t\gg 1000$ these states generate the noise looking like short-period oscillations with the wavelength of the order of the radial grid step $h$. This problem requires the circumspective approach because some oscillations in $\gamma(E)$ appear also in ——— calculations \cite{Kheifets2006}. But all indicates that oscillations in our results are the numerical artifact and come from the bound states destruction: its wavelenght is always of the order of the grid step for an any grid step value choice, ones appear when the $a(t)h$ becomes of the order of the typical bound state radius and spread toward to the increasing radius. For this reason we filtered the wave function after the evolution by an eliminating all components with wavelenghts less than $4h$. We should note that there are no oscillations in our calculations even without the filtration at the energy range where oscillations in C—— calculations \cite{Kheifets2006} appear. 

In the calculations of the Helium photoionization given below, we used the numerical scheme parameters (see \cite{Serov2008} for details) as follows: an angular basis parameter $l_{2\text{max}}$=13, a uniform radial grid having a number of the radial nodes $N_r$=500 and a size $\xi_{\text{max}}$=25., a complex scaling radius $\xi_{\text{sc}}$=22.5, a complex scaling angle $\theta_{\text{sc}}$=30$^\circ$, a grid expanding rate $\dot{a}_\infty$=0.1, the evolution was simulated up to the time $t_{\text{max}}=12800$. 
For other targets other radial grid parameters are used: $N_r$=500, $\xi_{\text{max}}$=50.,
$\xi_{\text{sc}}$=45., $\dot{a}_\infty$=0.05 for the H$^-$; $N_r$=1000, $\xi_{\text{max}}$=50., $\xi_{\text{sc}}$=40., $\dot{a}_\infty$=0.05 for the He in the excited $1s2s{\,}^1$S state; and  $N_r$=1400, $\xi_{\text{max}}$=70., $\xi_{\text{sc}}$=60., $\dot{a}_\infty$=1/30 for the He in the $1s3s{\,}^1$S state.

After the 3DCS has been calculated, it is necessary to obtain the Gaussian width parameter $\gamma$ from the 3DCS. The squared module of the correlation parameter $|a_g(E_1,E_1,\theta_{12})|^2$ may be expressed from the 3DCS through the Eq.(\ref{agau}) and, after that, fitted by (\ref{GaussianAppr}) using the least-squares method (LS) as in the theoretical work \cite{Kheifets2000}. An alternative approach is based on the fitting of the twofold differential cross section (2DCS) $\sigma^{(2)}(E_1,E_2,\theta_{12})=\frac{d^{2}\sigma}{dE_1d\theta_{12}}$ by the formula 
\begin{eqnarray}
\sigma^{(2)}(E_1,E_1,\theta_{12})
&\simeq&\frac{32\pi^2}{3}|A|^2\exp\left[-\frac{4\ln
2(\pi-\theta_{12})^2}{\gamma^2}\right]\cos^2\frac{\theta_{12}}{2}\label{apprsigma}
\end{eqnarray}
deduced from the Eq.(\ref{agau}) by integrating over all angles except the $\theta_{12}$. The analogous approach was used in the experimental work \cite{Huetz2000}. Since the correlation parameter $a_g(\theta_{12})$ can noticeably deviate from the Gaussian shape, the $\gamma$s calculated by two methods mentioned above are different. We will denote the Gaussian width parameter obtained by the fitting of $|a_g(\theta_{12})|^{2}$ as $\gamma(|a_g|^{2})$, and $\gamma$ obtained by the fitting of $\sigma^{(2)}(\theta_{12})$ as $\gamma(\sigma^{(2)})$.

\section{The photoinization of the Helium in the ground state}
\begin{figure}[h!]
\includegraphics[angle=-90,width=0.80\textwidth]{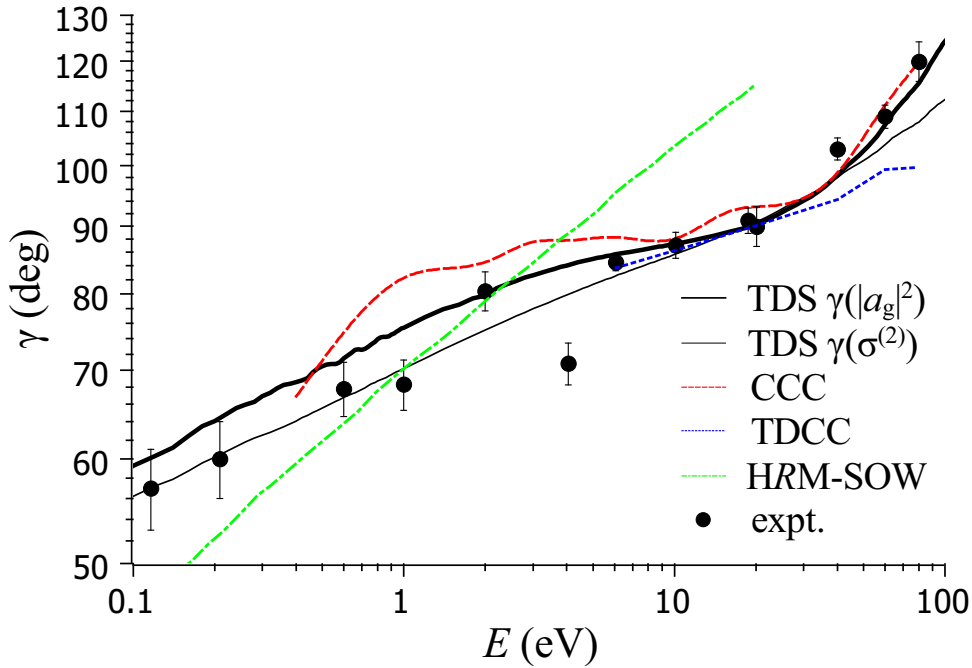}
\caption{The Gaussian width parameter $\gamma$ as a function of the full energy of ejected electrons $E$ for the photoionization of He: the TDS with the Gaussian fitting of $|a_g|^{2}$ (thick solid line) and the Gaussian fitting of $\sigma^{(2)}$ (thin solid line), CCC \cite{Kheifets2000,Kheifets2006} (dashed line), TDCC \cite{Colgan2006} (dotted line), H$R$M-SOW \cite{Kazansky1999} (dash-dotted line), and experimental data \cite{Dawber1995,Dorner1998,Malegat1997,Dawson2001,Huetz2000,Turri2002} (circles).\label{FIGgammaHe}}
\end{figure}
\begin{figure}[h!]
\includegraphics[angle=-90,width=0.50\textwidth]{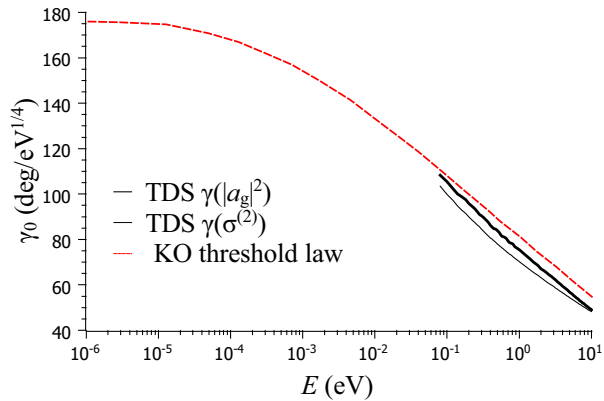}
\caption{The scaled width parameter $\gamma_0$ as a function of $E$ for He: the TDS with the Gaussian fitting of $|a_g|^{2}$ (thick solid line) and $\sigma^{(2)}$ (thin solid line), and  the Kasansky--Ostrovsky threshold law \cite{Ostrovsky1993} (dashed line).\label{FIGgamma0He}}
\end{figure}

In the Fig.\ref{FIGgammaHe} we show the Gaussian width $\gamma$ as a function of the full energy of ejected electrons $E$ for the photoionization of Helium in the ground state. It is obvious that our data coincides very well with the experiment in the whole range from 0.1 eV to 100 eV, except the point at 4 eV from \cite{Malegat1997}. The exact coincidence of our curve $\gamma(\sigma^{(2)})$ with the experimental points at 0.116 and 0.209 eV from \cite{Huetz2000} is the most remarkable. The curve $\gamma(|a_g|^{2})$ coincides with these points in rather less degree because $\gamma(\sigma^{(2)})$ has been got in the experiment \cite{Huetz2000}. Generally, the differences between $\gamma(|a_g|^{2})$ and $\gamma(\sigma^{(2)})$ may be the feature of the degree of the $a_g(\theta_{12})$ deviation from the Gaussian shape, although the $\gamma(|a_g|^{2})$ coincidence with the $\gamma(\sigma^{(2)})$ does not mean that $a_g(\theta_{12})$ is exactly Gaussian.
The Fig.\ref{FIGgammaHe} is given in the logarythmic scales at both axises, in which the power dependences like the Wannier law (\ref{WannierLaw}) should look like sloped straight lines. Indeed, we see that our plots are close to lines when $E$ is less than few eV. But the exponent is not equal to 1/4 at all! The approximation of $\gamma(\sigma^{(2)})$ curve at the $E$ range from 0.1 to 2 eV using the power law of the general form
\begin{eqnarray}
\gamma=\tilde{\gamma}_0 E^{s} \label{SerovLaw}
\end{eqnarray}
through the least-squares approach yields the exponent $s$=0.097 and the proportionality constant $\tilde{\gamma}_0$=70.2$^\circ$ eV$^{-s}$. Such a significant deviance from the Wannier threshold law often used for the interpretation of experimental and theoretical data seems to be discouraging. However, Kazansky and Ostrovsky \cite{Ostrovsky1993} show that the Wannier threshold law strongly modifies when the electrons deceleration by the nucleus field is taken into consideration because the electrons velocity is non-equal to zero even at zero total energy. In the Fig.\ref{FIGgamma0He} we show the scaled
width parameter $\gamma_0(E)=\gamma(E)/E^{1/4}$ in compare with the approximate curve obtained in \cite{Ostrovsky1993}, which we called the Kasansky--Ostrovsky threshold law.  

\section{The photoionization of the targets with the strongly asymmetrical initial state configuration}
\begin{figure}[h!]
\includegraphics[angle=-90,width=0.50\textwidth]{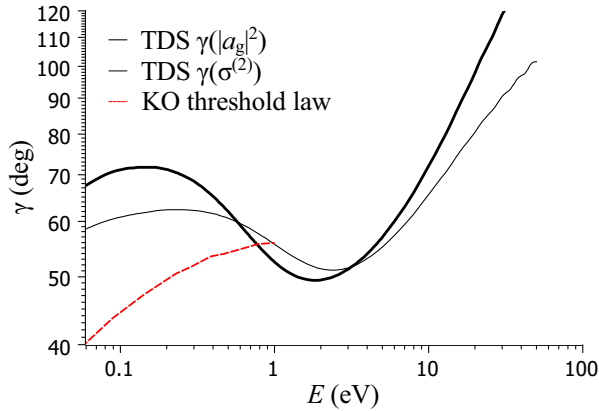}
\caption{The Gaussian width $\gamma$ as a function of $E$ for the photoionization of H$^-$: the TDS with the Gaussian fitting of $|a_g|^{2}$ (thick solid line) and $\sigma^{(2)}$ (thin solid line), and the Kasansky--Ostrovsky threshold law \cite{Ostrovsky1993} (dashed line).\label{FIGgammaHminus}}
\end{figure}
In our previous work \cite{Serov2008} we did a comparison of $\gamma(E)$ for various Helium-like ions and found that for the negative Hydrogen ion H$^-$ it starts to increase at energies below 2.5 eV unlike other considered targets. Here we obtained the results at the energies down to 0.06 eV. It is clear from the Fig.\ref{FIGgammaHminus} that $\gamma(\sigma^{(2)})$ increases at the energy decreasing in the range from 2.6 eV to 0.23 eV. At energies below 0.09 eV $\gamma(\sigma^{(2)})$ is a power function of the energy with the exponent $s$=0.083 and $\tilde{\gamma}_0$=74$^\circ$ eV$^{-s}$, though this range is too small for such rigorous conclusions. 
\begin{figure}[h!]
\includegraphics[angle=-90,width=0.50\textwidth]{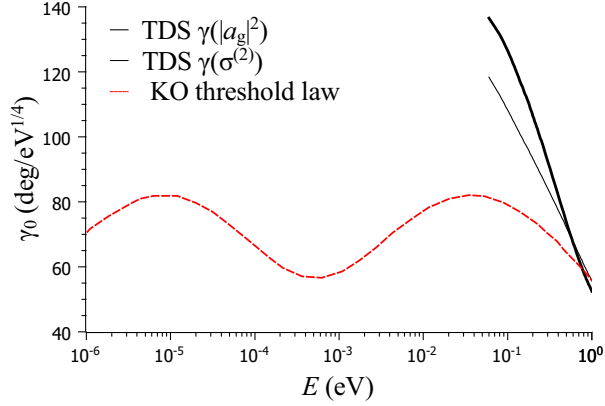}
\caption{The scaled width parameter $\gamma_0$ as a function of $E$ for H$^-$: the TDS with the Gaussian fitting of $|a_g|^{2}$ (thick solid line) and $\sigma^{(2)}$ (thin solid line), and the Kasansky--Ostrovsky threshold law \cite{Ostrovsky1993} (dashed line).\label{FIGgamma0Hminus}}
\end{figure}
It is obvious from the Fig.\ref{FIGgammaHminus} that there is a strong distinction of our results from the Kasansky--Ostrovsky threshold law for the nuclear charge $Z=1$ \cite{Ostrovsky1993}, unlike the Helium case (Fig.\ref{FIGgamma0He}). We should note that the $\gamma(E)$ dependence obtained by Kasansky and Ostrovsky is monotonous (Fig.\ref{FIGgammaHminus}) despite the $\gamma_0(E)$ dependence oscillating.
Our hypothesis is that such distinction of the results for H$^-$ from the results for the Helium and other Helium-like ions \cite{Serov2008} comes from the fact that the H$^-$ bound state configuration is strongly different from the one of the Helium ground state. Indeed, when $r_{1,2}\to\infty$ the H$^-$ bound state wave function has an asymptotic
\[ \Phi(\mathbf{r}_1,\mathbf{r}_2) \sim e^{-r_1}\frac{e^{-0.235r_2}}{r_2}+e^{-r_2}\frac{e^{-0.235r_1}}{r_1}. \]
The H$^-$ is a deuteron-like weakly bound system consisting from the Hydrogen atom and the electron located the most time outside the region where the attracting potential acts. We performed here the calculations for the another targets with the strongly asymmetrical initial state configuration: Helium atoms in the excited states $2s{\,}^1$S and $3s{\,}^1$S. 

\begin{figure}[h!]
\includegraphics[angle=-90,width=0.50\textwidth]{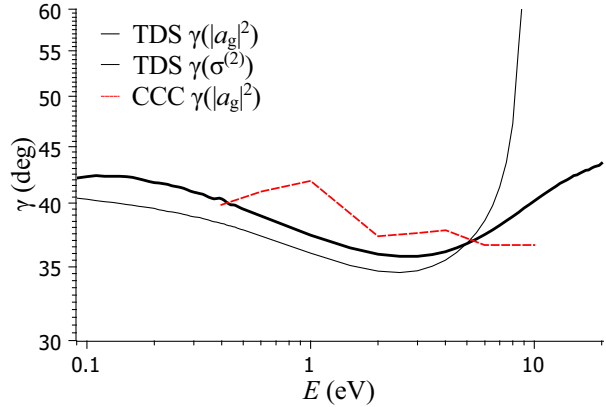}
\caption{The Gaussian width $\gamma$ as a function of $E$ for the photoionization of He($2s{\,}^1$S): the TDS with the Gaussian fitting of $|a_g|^{2}$ (thick solid line) and $\sigma^{(2)}$ (thin solid line), and CCC results \cite{Kheifets2006}.\label{FIGgammaHe2s}}
\end{figure}
In the Fig.\ref{FIGgammaHe2s} we show $\gamma$ as a function of $E$ for the photoionization of Helium in $2s{\,}^1$S metastable state. Our results do not deviate much from the CCC results \cite{Kheifets2006} by the magnitude but deviate strongly by the behavior because there are no oscillations in our results. When the energy decreases $\gamma$ decreases at first and then begin to increase (for $\gamma(\sigma^{(2)})$ it happens at $E$=2.5 eV) as well as for H$^-$. Surprisingly, at $E$>5 eV the difference between $\gamma(|a_g|^{2})$ and $\gamma(\sigma^{(2)})$ became enormous. 
\begin{figure}[h!]
\includegraphics[angle=-90,width=0.50\textwidth]{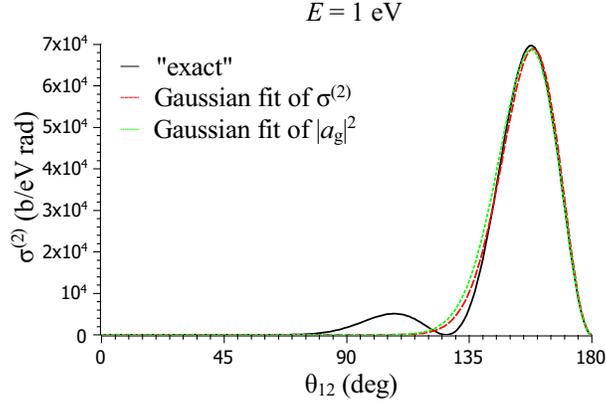}\\
(a)\\
\includegraphics[angle=-90,width=0.50\textwidth]{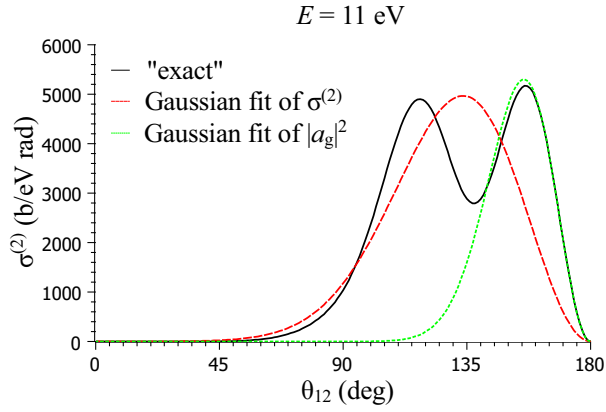}\\
(b)\\
\includegraphics[angle=-90,width=0.50\textwidth]{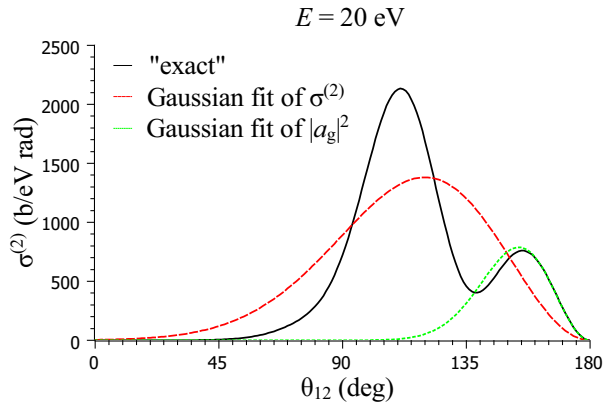}\\
(c)
\caption{The 2DCS as a function of the interelectron angle $\theta_{12}$ for the photoionization of He($2s{\,}^1$S) for a) $E$=1 eV; b) $E$=11 eV; and c) $E$=20 eV: ``exact'' TDS results (solid line), Gaussian fitting of $\sigma^{(2)}$ (dashed line) and the Gaussian fitting of $|a_g|^{2}$ (dotted line).\label{FIGsigma2He2s}}
\end{figure}
The origin for this is clear from the Fig.\ref{FIGsigma2He2s}. The correlation coefficient has a strongly non-Gaussian shape with the two peaks even at low energy $E$=1 eV. When the energy decreases the secondary peak declines and therefore the distribution turns to Gaussian as follows from the Wannier's theory. When the energy increases the secondary peak grows, and becomes more than the primary one at $E$=11.3 eV as it is seen from the Fig.\ref{FIGsigma2He2s}b,c. Then the Gaussian fitting becomes non-applicable of course.  

\begin{figure}[h!]
\includegraphics[angle=-90,width=0.50\textwidth]{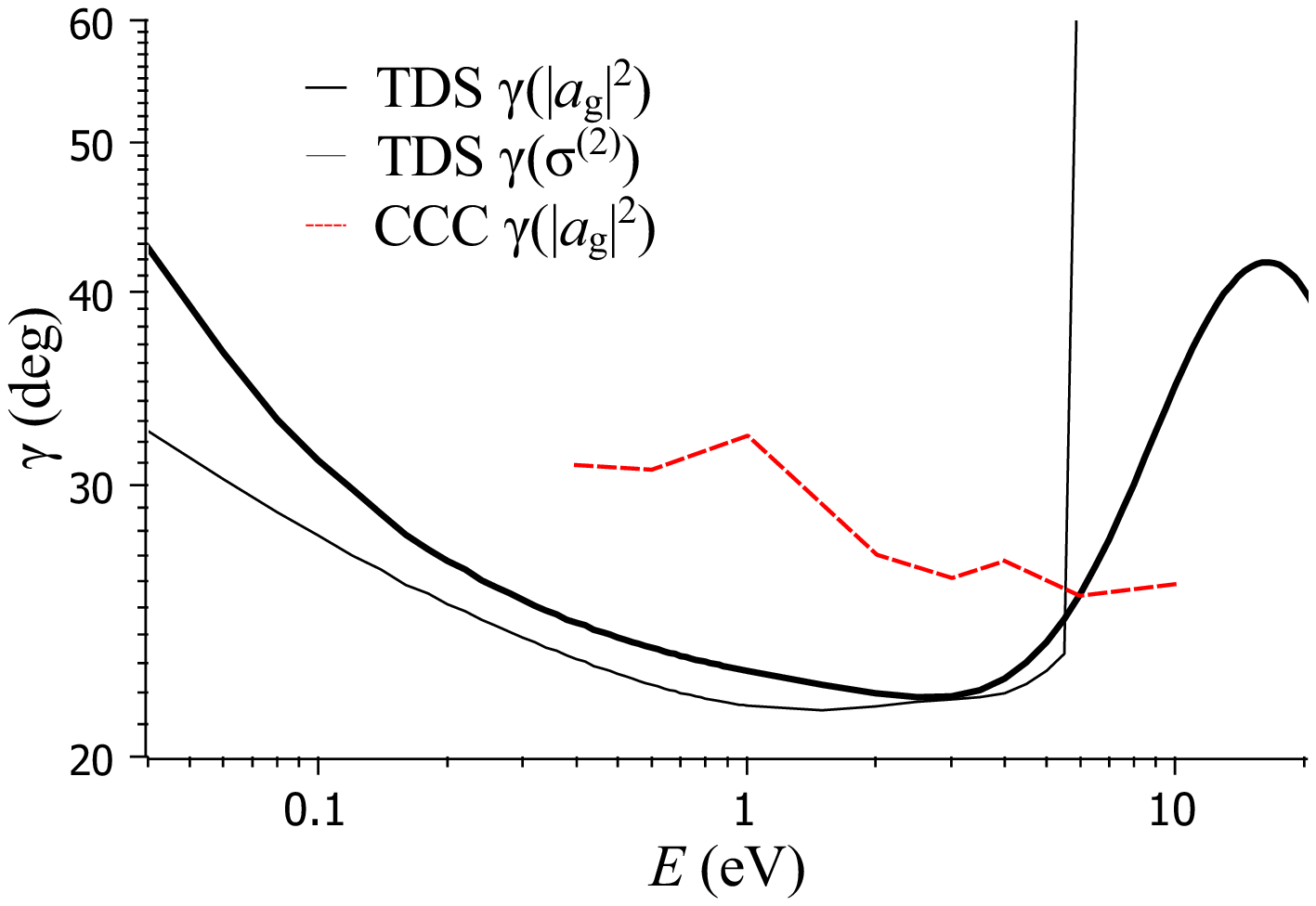}
\caption{The Gaussian width $\gamma$ as a function of $E$ for the photoionization of He($3s{\,}^1$S): the TDS with the Gaussian fitting of $|a_g|^{2}$ (thick solid line) and $\sigma^{(2)}$ (thin solid line), and CCC results \cite{Kheifets2006}.\label{FIGgammaHe3s}.}
\end{figure}
In the Fig.\ref{FIGgammaHe3s} we show $\gamma$ as a function of $E$ for the photoionization of the Helium in the $3s{\,}^1$S state. Our results are strongly different from the CCC results \cite{Kheifets2006} both by the magnitude and by the behavior. The general curve shape is similar to the one for $2s{\,}^1$S, but $\gamma(\sigma^{(2)})$ reaches the local maximum at $E$=1.5 eV.
\begin{figure}[h!]
\includegraphics[angle=-90,width=0.50\textwidth]{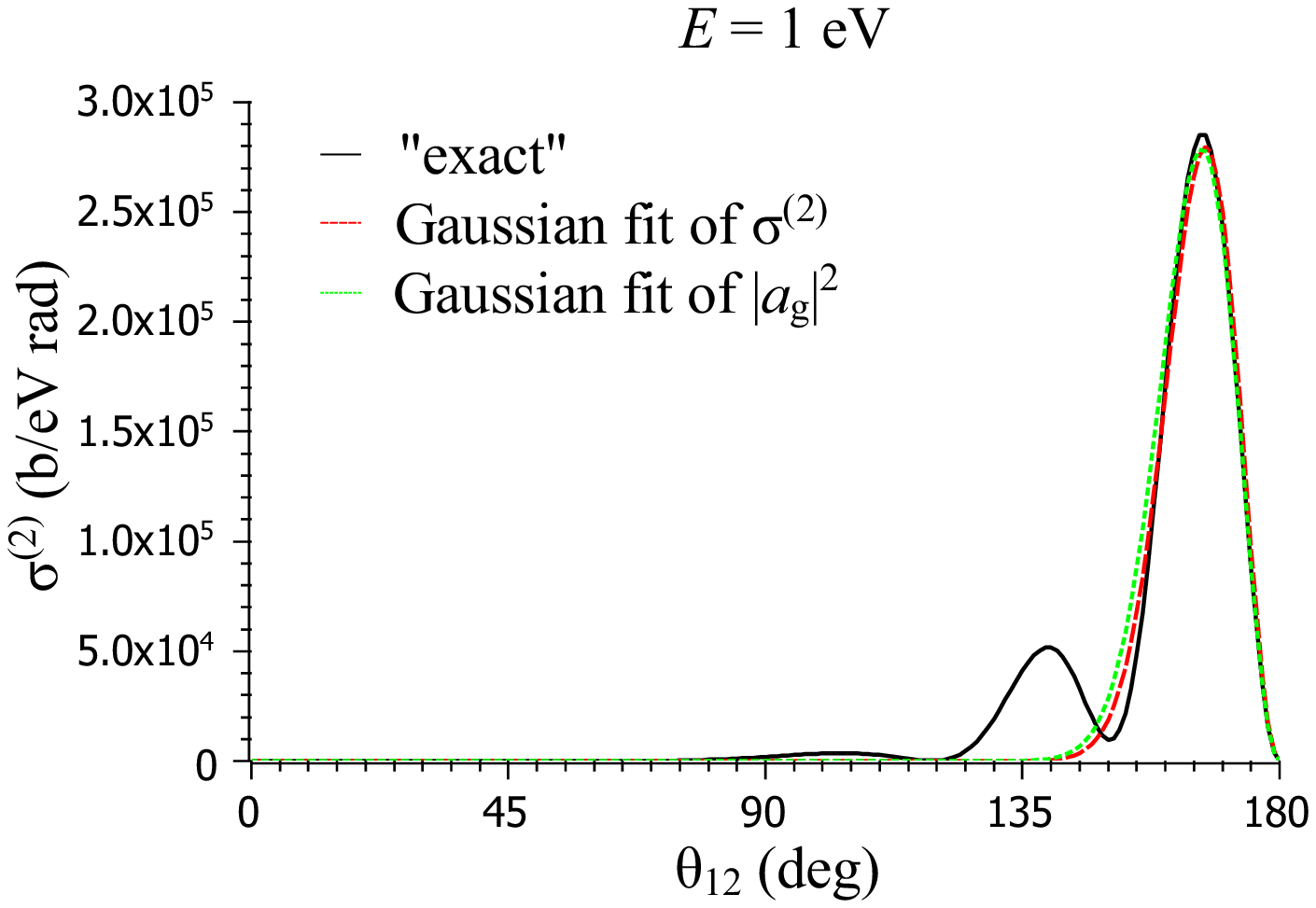}\\
(a)\\
\includegraphics[angle=-90,width=0.50\textwidth]{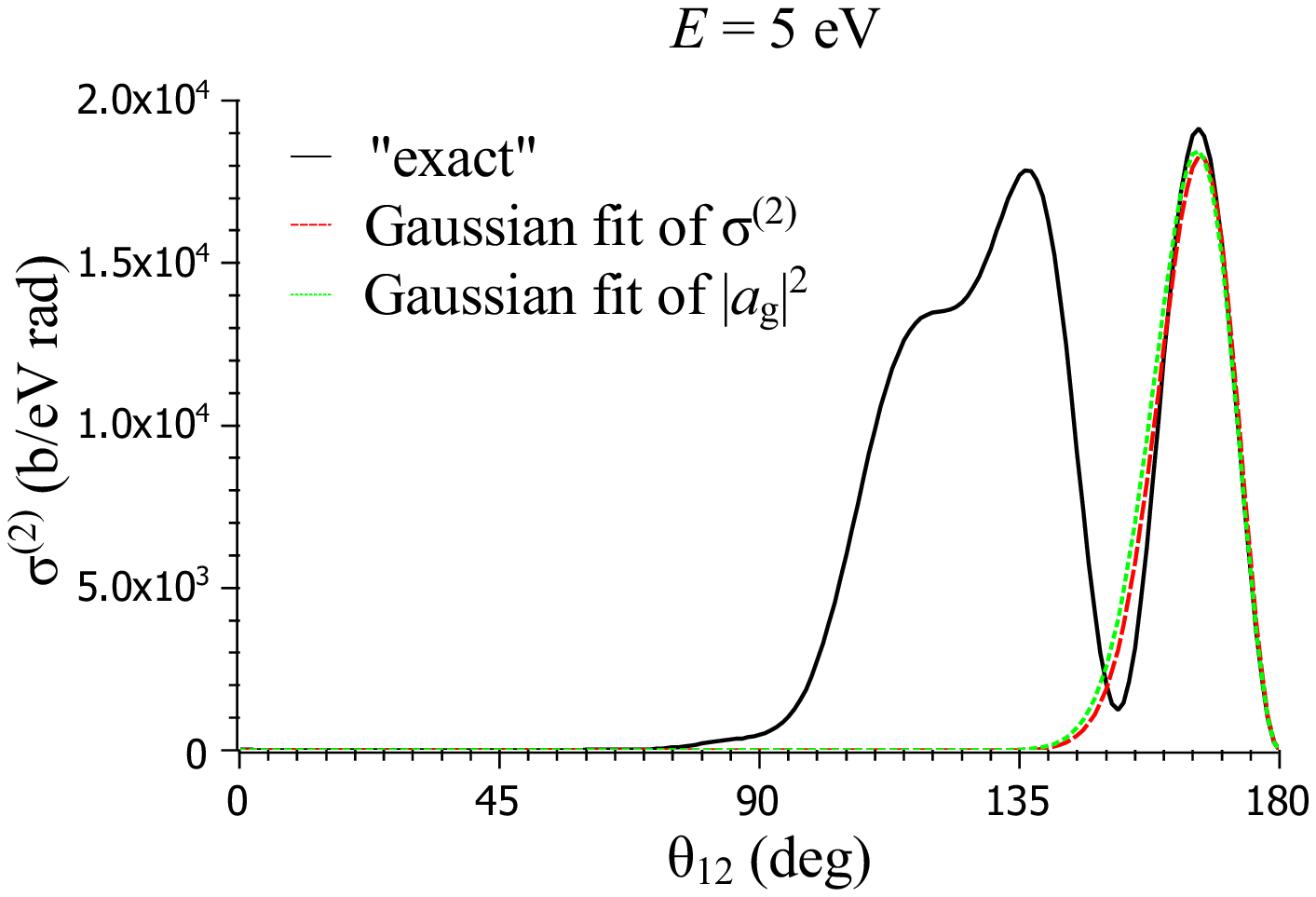}\\
(b)\\
\includegraphics[angle=-90,width=0.50\textwidth]{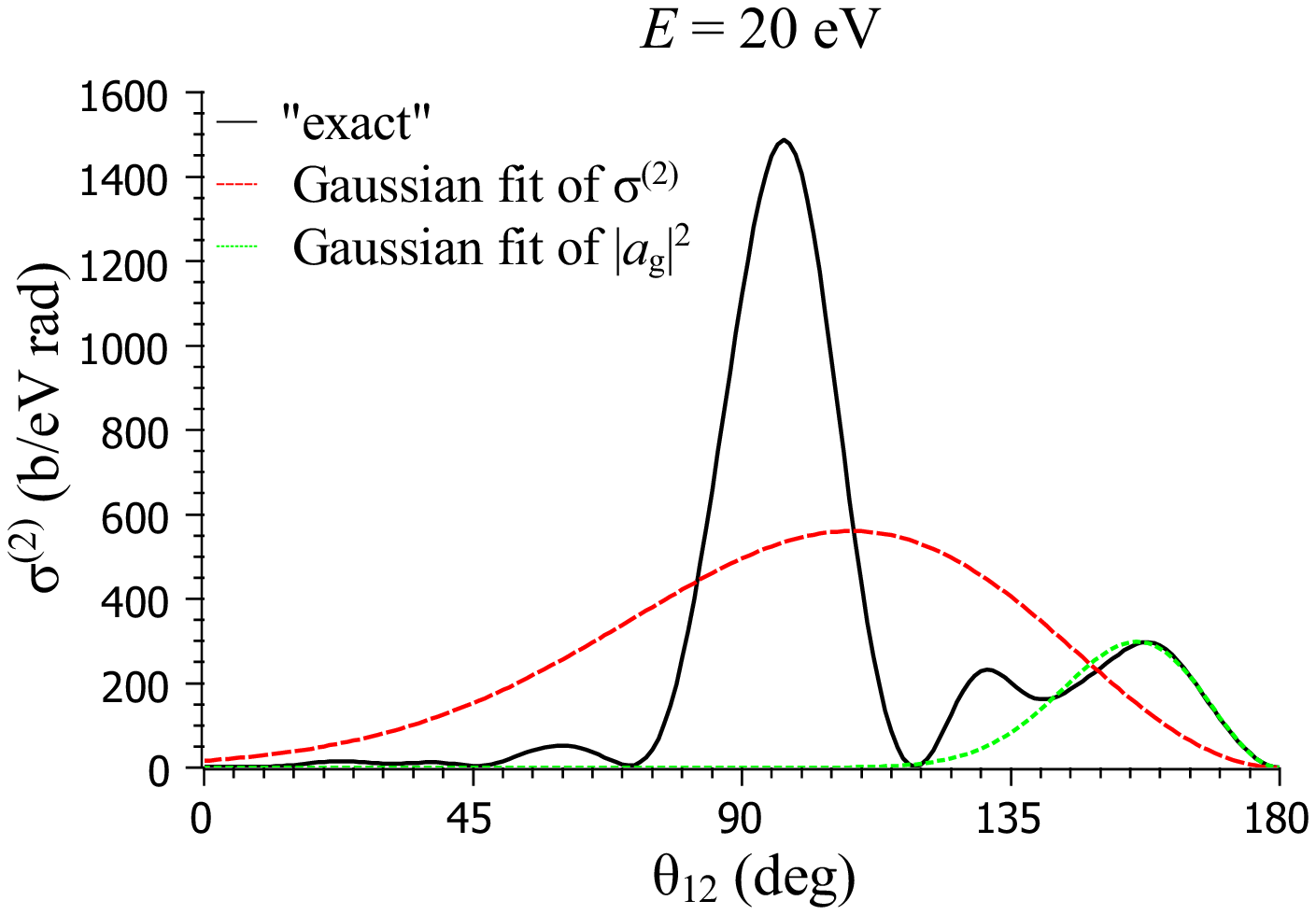}\\
(c)
\caption{The 2DCS as a function of the interelectron angle $\theta_{12}$ for the photoionization of He($3s{\,}^1$S) for a) $E$=1 eV; b) $E$=5 eV; and c) $E$=20 eV: ``exact'' TDS results (solid line), Gaussian fitting of $\sigma^{(2)}$ (dashed line) and the Gaussian fitting of $|a_g|^{2}$ (dotted line).\label{FIGsigma2He3s}}
\end{figure}
It is seen from the Fig.\ref{FIGsigma2He3s} that the correlation parameter is strongly non-Gaussian as well as for $2s{\,}^1$S, but has three peaks at low energies. When the energy increases two lesser peaks grow and merge, and the $\sigma^{(2)}(\theta_{12})$ view becomes rather complicated at large energies. One can see that the number of considered targets $\sigma^{(2)}(\theta_{12})$ peaks at low energies is equal to the number of the peaks of the target ``outer'' electron density dependence on the radius in the initial state. Particularly, it is equal to one for H$^-$, two for He($2s{\,}^1$S) and three for He($3s{\,}^1$S). It seems not to be an accidental coincidence though we still can not propose an exact explanation of this effect. If this dependence will be confirmed it will reveal the new abilities for the qualitative experimenatal analysis of the target electron structure using the over-threshold double photoionization by analogy with the $(e,2e)$ spectroscopy \cite{Popov1994}. 

When we have already established that the local minimum appearance in the $\gamma(E)$ is common for all considered targets with the asymmetrical initial state configuration we can explore the origin of this effect. For all targets the local minima of the $\gamma$ are observed at the energies not depending direct proportionally on the target first ionization potential $I_1$, but of the order of $I_1$. We should note that despite of the absence of the local minimum on the plot for the Helium in the ground state (Fig.\ref{FIGgammaHe}) there is a curve bend at $E$=20 eV approximately that indicates to some change taking place at the energy of the order of $I_1$. In the non-sequential double ionization considered here one of the electrons is ionized at first by the photon impact, and the second electron might be ejected through the a sudden change of the atomic potential (so called the shake-off mechanism), or through the first electron impact (so called the final state scattering). If the initial state configuration is strongly asymmetrical then the ``inner'' electron momentum density is much broader than the one for the ``outer'' electron, and the single ionization cross section dependence on the ejection energy for the ``inner'' electron decreases much slower than for the ``outer'' one. It means that the process when the ``outer'' electron is ionized first may give a significant contribution only when $E$ is of order of the ``outer'' electron binding energy, i.e. $I_1$. From the other hand, it is obvious that the shake-off mechanism may give a significant contribution only in the case when the first ionized electron velocity is much larger than the velocity distribution width for the second one. Indeed, when the velocity of the first ionized electron is comparable to the velocity of the bound one, then the potential affecting to the second electron changes slowly, and the second electron adiabatically turns to the bound state with the same symmetry as its initial state due to the well-known theorem, therefore its ionization can not proceed. Therefore, when the ``outer'' electron is ionized at first then the shake-off mechanism is impossible. But the possibility of the final state scattering is also extremely small if the ``outer'' electron is ionized at first because it can not have an energy enough for the ejection of the strongly bounded ``inner'' electron. Consequently, the double ionization is possible if only the ``inner'' electron is ionized at first. In this case the shake-off mechanism is significant only at emission energies much larger than the first ionization potential. From the comparing of $\gamma$ for the double photoionization with $\gamma$ for $(e,2e)$ in \cite{Kheifets2006} it is clear that the angular interelectron correlations are stronger for the shake-off process then for the final state scattering. Summarizing all written above we can conclude that $\gamma$ decreases with the energy decreasing for both processes taken individually, and the $\gamma$ increasing at low energies in the Figs. \ref{FIGgammaHminus}, \ref{FIGsigma2He2s} and \ref{FIGsigma2He3s} results from the ``switching off'' of the shake-off process. The distinction from the He in the ground state is just the sharpness of this switching off because of the narrow momentum spectrum of the ``outer'' electron. It may be predicted, that in the double photoionization of targets with $I_1$ much less than $I_2$ (like alkali metal atoms), $\gamma(E)$ should behave similarly the one for H$^-$ on Fig. \ref{FIGgammaHminus}.

\section{Conclusion}
We calculated {\it ab initio} the Gaussian width $\gamma$ dependence on the electrons energy $E$ in the double photoionization of the negative Hydrogen ion H$^-$, the Helium in the ground $1s^2$ state and $2s{\,}^1$S and $3s{\,}^1$S excited states. For the He($1s^2$) photoionization our results are in the excellent agreement with experimental data but indicate that the well-known Wannier threshold law $\gamma\propto E^{1/4}$ is not correct at energies attainable in modern experiments. It is shown that the $\gamma$ dependence on the energy is much better described by the curve obtained by Kasansky and Ostrovsky when the electrons interaction with the nucleus was taken into cosideration. 
We have shown that for all considered targets with the strongly asymmetrical initial state configuration $\gamma(E)$ has the region of the decreasing when $E$ has the order of the magnitude of the first ionization potential $I_1$. We suppose that this effect is connected with the rapid alternation from the dominating of the shake-off mechanism to the less-correlated final state scattering. Also, it was demonstrated that the cross section dependence on the interelectron energy for these targets is strongly non-Gaussian at low energies and has a number of the peaks equal to the number of the initial state radial nodes, and it reveals the new abilities for the targets electron structure qualitative analysis. 

\acknowledgments The authors are grateful to Prof. V. Derbov
for help and discussions. This work was supported by President of Russian
Federation Grant No. MK-2344.2010.2 and Grant No. RFBR
08-01-00604a.

\end{document}